\documentstyle[12pt,psfig]{article}
\def\baselinestretch{1.0}
\begin{document}
\pagenumbering{arabic}
\begin{center}
{\Large \bf Bound states of anti-nucleons in finite nuclei} \\
\vspace{0.8cm}
{G.~Mao, H.~St\"{o}cker and W.~Greiner}  \\ 
{\it Institut f\"{u}r Theoretische Physik der
J. W. Goethe-Universit\"{a}t \\
Postfach 11 19 32,  D-60054 Frankfurt am Main, Germany}
\end{center}
\vspace{1.2cm}
\begin{abstract}
\begin{sloppypar}
We study the bound states of anti-nucleons emerging from the lower continuum
in finite nuclei within
the relativistic Hartree approach including the contributions of 
the Dirac sea to the source terms of the meson fields. The Dirac equation
is reduced to two Schr\"{o}dinger-equivalent equations for
the nucleon and the anti-nucleon respectively. These two equations
 are solved simultaneously in an iteration procedure. 
 Numerical results show that the bound levels of 
anti-nucleons vary  drastically when the vacuum contributions are taken into 
account.

\end{sloppypar}
\bigskip
\noindent {\bf PACS} number(s): 21.10.-k; 21.60.-n; 03.65.Pm             
\end{abstract}
\newcounter{cms}
\setlength{\unitlength}{1mm}
\newpage
 \begin{sloppypar}
In spite of the great successes of the relativistic mean field (RMF) theory
\cite{Ser86,Rei86,Gam90,Ren95}
and the relativistic Hartree approach (RHA) \cite{Hor84,Fur89} in describing
the ground states of nuclei, the arguments of introduction of strong Lorentz
scalar ($S$) and time-component Lorentz vector ($V$) potential in the Dirac 
equation are largely indirect. So far, no evidence from experiments ensures
the physical necessity. One usually compares the theoretical 
predictions only with the
experimental data of the nucleon sector (i.e., the shell-model states),
which is subject to a
relatively small quantity stemming from the cancellation of two potentials
$S+V$ ($V$ is positive, $S$ is negative.). While the dynamical content of the
Dirac picture certainly lies with both the nucleon and the anti-nucleon
sector, the study of the anti-nucleon sector enjoys an additional bonus:
it provides us with a chance to determine the individual $S$ and $V$! 
Due to the
{\em G-parity}, the vector potential changes its sign in the anti-nucleon
sector. The bound states of anti-nucleons are sensitive to the sum of 
the scalar and vector field $S-V$. Combining with the information from the
nucleon sector, one may fix the individual values of the scalar and
vector field.

The study of the anti-nucleon bound states is extremely interesting for 
modern nuclear physics.
If the potential of anti-nucleons is much weaker than what one 
expects or predicts by means of the  RMF/RHA models, 
that is, the strong scalar and vector field are not necessary,
one may question the
validity of the models since some important physical ingredients,
such as quantum corrections, correlation effects, three-body forces
 et al., are still missing
in these phenomenological approaches. One may think about constituting a more
elaborate model. Alternatively,
if a deep potential of 
 anti-nucleons is indeed observed, that is, the strong scalar and vector
potential are realistic, an interesting phenomenon is that at certain density
the energy of anti-nucleons may turn out to be larger than the free nucleon
 mass, the system becomes unstable with respect to the 
nucleon--anti-nucleon pair creation \cite{Mis93}. On the other hand, as 
pointed out in Ref. \cite{Gre95}, in high-energy relativistic heavy-ion
collisions, the nucleons may be emitted from the deep bound states emerging 
from the Dirac sea due to dynamics. These can create a great number of 
anti-nucleons in bound states. Such collective 
creation processes of anti-matter clusters have a large probability for the
production of anti-nuclei, -- and analogously also for multi-$\Lambda$,
multi-$\bar{\lambda}$ nuclei. These open two fascinating directions to 
extend the periodic system, i.e., to extend into the anti-nucleon sector and
into the multi-strangeness dimension, in addition to the islands of super-heavy
nuclei. In order to reach the quantitative study of the above theoretical 
conjecture, a prerequisite is to know the exact potential depth of the bound 
states of anti-nucleons. Up to now, no answers from experimental side or
theoretical side are available.

This is the aim of the present work. We study the problem within the
relativistic Hartree approach including the vacuum contributions. The starting 
point is the following effective Lagrangian for nucleons interacting through
the exchange of mesons \cite{Ser86,Rei86,Gam90}
   \begin{eqnarray}
{\cal L}&=&\bar{\psi}[i\gamma_{\mu}\partial^{\mu}-M_{N}]\psi
   + \frac{1}{2}
\partial_{\mu}\sigma\partial^{\mu}\sigma-U(\sigma)
 -\frac{1}{4}\omega_{\mu\nu}\omega^{\mu\nu} \nonumber \\
&& + \frac{1}{2}m_{\omega}^{2}\omega_{\mu}\omega^{\mu}
 - \frac{1}{4} {\bf R}_{\mu\nu}{\bf R}^{\mu\nu}
 +\frac{1}{2}m_{\rho}^{2}{\bf R}_{\mu} \cdot {\bf R}^{\mu}
- \frac{1}{4}  A_{\mu\nu} A^{\mu\nu} \nonumber \\
&& + {\rm g}_{\sigma}\bar{\psi}\psi\sigma
      - {\rm g}_{\omega}\bar{\psi}\gamma_{\mu}\psi\omega^{\mu}
 - \frac{1}{2}{\rm g}_{\rho}\bar{\psi}\gamma_{\mu}\mbox{\boldmath $\tau$}
\cdot \psi {\bf R}^{\mu} - \frac{1}{2} e \bar{\psi}(1+\tau_{0})\gamma_{\mu}
 \psi A^{\mu},
    \end{eqnarray}
here U($\sigma$) is the self-interaction part of the scalar field
\cite{Bog83}
\begin{equation}
  U(\sigma)=
   \frac{1}{2}m_{\sigma}^{2}\sigma^{2}+\frac{1}{3!}b
\sigma^{3}+\frac{1}{4!}c\sigma^{4}.
\end{equation}
Based on this Lagrangian, we have developed
 a relativistic model describing the bound states of both nucleons and
 anti-nucleons
in finite nuclei \cite{MaoPre}. Instead of directly searching for 
two solutions of the Dirac equation in a finite many-body system,
 we reduce the Dirac equation
to two Schr\"{o}dinger-equivalent equations for the nucleon and
the anti-nucleon respectively.  These two equations can be solved
simultaneously with the numerical technique of the relativistic mean-field
theory for finite nuclei.
The contributions of the Dirac sea to the source terms of the
meson fields are evaluated by means of the derivative expansion \cite{Ait84}
up to the leading derivative order for the one-meson loop and one-nucleon loop.
Thus, the wave functions of anti-nucleons, which are used to calculate the
single-particle energies,  are not involved in evaluating the
vacuum contributions to the scalar and baryon density which are, in turn, 
expressed by means of the scalar and vector field as well as their derivative
terms \cite{MaoPre}. The Schr\"{o}dinger-equivalent equation of the nucleon
and the equations of
motion of mesons (containing the densities contributed from the vacuum) 
are solved within a self-consistent iteration procedure
\cite{Rei86}. Then, the equation of the anti-nucleon is solved with the
known mean fields to obtain the wave functions and the single-particle energies 
of anti-nucleons. The space of anti-nucleons are truncated by the specified 
principal and angular quantum numbers $n$ and $j$ with the guarantee  that
the calculated single-particle energies of anti-nucleons are converged when the
truncated space is extended.
We find that the 
results are insensitive to the exact values of $n$ and $j$ provided large 
enough numbers are given. 
We have used $n=4$, $j=9$ for $^{16}{\rm O}$; 
$n=5$, $j=11$ for $^{40}{\rm Ca}$;
and $n=9$, $j=19$ for $^{208}{\rm Pb}$.                       

In the previous RHA calculations for the bound states of nucleons
\cite{Hor84,Fur89}, the parameters of the model are fitted to the saturation
properties of nuclear matter as well as the $rms$ charge radius in 
$^{40}{\rm Ca}$. The {\em best-fit} routine within the RHA to the properties of 
spherical nuclei has not been performed yet. Thus, we first fit the parameters
of Eqs. (1)  and (2) within the RHA to the empirical data of binding energy,
surface thickness and diffraction radius of 
eight spherical nuclei $^{16}{\rm O}$, $^{40}{\rm Ca}$, 
$^{48}{\rm Ca}$, $^{58}{\rm Ni}$, $^{90}{\rm Zr}$, $^{116}{\rm Sn}$,
$^{124}{\rm Sn}$, and $^{208}{\rm Pb}$
as has been done in
Ref. \cite{Rei86} for the RMF model. We distinguish two different cases with
(RHA1) and without (RHA0) nonlinear self-interaction of the scalar field.
The obtained parameters and the corresponding saturation properties are 
given in Table I. For the sake of comparison, 
two sets of the linear (LIN) and nonlinear (NL1)
RMF parameters from Ref. \cite{Rei86} 
are also presented. One can see that the RHA gives a larger
effective nucleon mass than the RMF does, which is mainly caused by the
feedback of the vacuum to the meson fields, as can be seen from Eqs. (71) $\sim$
(74) of Ref. \cite{MaoPre}. When the effective nucleon mass decreases, the
scalar density originated from the Dirac sea $\rho_{S}^{sea}$ increases.  
It is negative
and cancels part of the scalar density contributed from the valence nucleons 
$\rho_{S}^{val}$, which causes the effective nucleon mass to increase again.
At the end, it reaches a balance value. 
In the fitting procedure, we have tried different initial values giving 
smaller effective nucleon mass. After running the code many times, all of them
slowly converge to a large $m^{*}$. 
The larger effective nucleon mass  
 explains why a larger $\chi^{2}$ value is
obtained for the RHA1 compared to the NL1. 
If one uses the current nonlinear RMF/RHA
models to fit the ground-state properties of spherical nuclei, an 
effective nucleon mass around 0.6 is preferred. The situation, however, might
be changed when other physical ingredients, e.g., tensor couplings,
correlation effects,
three-body forces, are taken into account, which warrants further investigation.
On the other hand, in the case of linear model, the RHA0 gives a better fit 
than the LIN does.    
This is mainly due to the vacuum contributions which
improve the theoretical results of the surface thickness substantially, 
and finally
improve the total $\chi^{2}$ value. An interesting quantity is the shell
fluctuation which can be best expressed via the charge density 
in $^{208}{\rm Pb}$ as
 \begin{equation}
\delta\rho =\rho_{C}(1.8\;\; {\rm fm}) - \rho_{C}(0.0\;\; {\rm fm}).
 \end{equation}
The empirical value is $-0.0023$ ${\rm fm}^{-3}$ \cite{Rei86}, 
which is nicely reproduced
in the RHA (see Table I) while the RMF overestimates $\delta\rho$ by
a factor of 3, sharing the same disease with the non-relativistic mean field
theory \cite{Fri86}.

In Table II we present the results of both the proton and the anti-proton
spectra of $^{16}{\rm O}$, $^{40}{\rm Ca}$ and $^{208}{\rm Pb}$. The 
binding energy per nucleon and the {\em rms} charge radius are given too.   
The numerical calculations are performed within two frameworks, i.e., the RHA
including the contributions of the Dirac sea to the source terms
of the meson fields and the RMF taking into account
 only the valence nucleons as the 
meson-field sources. The experimental data are taken from Ref. \cite{Mat65}.
From the table one can see that all four sets of parameters can reproduce
the empirical values of the binding energies, the {\em rms} charge radii
and the single-particle energies of protons fairly well. 
For the $E/A$
and the $r_{ch}$, the agreement between the theoretical predictions and the
experimental data are improved from the LIN to the RHA0, RHA1 and NL1 set
of parameters. For the spectra of protons, due to large
error bars, it seems to be difficult to queue up 
the different sets of parameters. However,
because of the large effective nucleon mass, 
the RHA has a smaller spin-orbit splitting (see $1p_{1/2}$ and $1p_{3/2}$ 
state) compared to the RMF. This situation can be improved through introducing 
a tensor coupling for the $\omega$ meson \cite{Rei86} 
which will be investigated in the future studies. 
For the anti-nucleon sector, no experimental
data are available. In all four cases, the potential of anti-protons 
is much deeper than the potential of protons.
On the other hand, 
one can notice the drastic difference between the RHA and
the RMF calculations -- the single-particle energies of anti-protons
calculated from the RHA
are about half of that from the RMF,  
exhibiting the importance of taking into account the
Dirac sea effects. It demonstrates that the anti-nucleon spectra deserve
a sensitive probe to the effective interactions. 
The spin-orbit splitting of the anti-nucleon sector is so
small that one nearly can not distinguish the $1\bar{p}_{1/2}$ and the 
$1\bar{p}_{3/2}$ state. This is because the spin-orbit potential is related to
$d(S+V)/dr$ in the anti-nucleon sector and two fields cancel each 
other to a large extent. 
Nevertheless, the space between the $1\bar{s}$ and the $1\bar{p}$ state is
still evident, especially for lighter nuclei. This might be helpful to 
separate the process of knocking out a $1\bar{s}_{1/2}$ 
nucleon from the background -- a promising way to measure the potential of
the anti-nucleon in laboratory.

In summary, we have proposed to study the bound states of anti-nucleons
in finite nuclei which will provide us with a chance to judge the physical 
necessity of introducing
strong scalar and vector potential in the Dirac picture. Due to
the feedback of the vacuum to the meson fields, the scalar and vector fields 
decrease in the RHA. Numerical calculations show that the single-particle
energies of anti-nucleons change drastically in the RMF and the
RHA approach for  different sets of parameters, while the single-particle 
energies of nucleons remain in a reasonable range. It is very
important to have experimental data to check the theoretical predicted bound
levels of anti-nucleons. If the Dirac picture with the 
large potentials is valid
for nucleon-nucleus interactions, a fascinating direction of future studies
is to investigate the vacuum correlation and the collective production of 
 anti-nuclei in relativistic heavy-ion collisions. Experimental efforts
in this direction are presently underway \cite{Ars99}.

The authors thank P.-G.~Reinhard, Zhongzhou~Ren, J.~Schaffner-Bielich
 and C.~Beckmann for stimulating discussions.
 G.~Mao gratefully acknowledges the Alexander von Humboldt-Stiftung for 
 financial support and  the people at the
Institut f\"{u}r
Theoretische Physik der J.~W.~Goethe Universit\"{a}t for their hospitality.
This work was supported by DFG-Graduiertenkolleg Theoretische und
Experimentelle
Schwerionenphysik, GSI, BMBF, DFG and A.v.Humboldt-Stiftung.                             

 \end{sloppypar}

\newpage
 \def\baselinestretch{1.0}
\begin{table}
\caption{Parameters of the RMF and the RHA models as well as the corresponding 
saturation properties. The results of shell fluctuation and 
the $\chi^{2}$ values of different sets of parameters are
also presented.} \vspace{0.5cm}
\begin{tabular}{lcccc}
\hline
\hline
 & \multicolumn{2}{c}{RMF} & \multicolumn{2}{c}{RHA}    \\
 & LIN & NL1 & RHA0 & RHA1  \\
 \hline
$M_{N}$ (MeV)        &  938.000    &  938.000 &  938.000 &   938.000   \\
$m_{\sigma}$ (MeV)   &  615.000    &  492.250 &  615.000 &   458.000   \\
$m_{\omega}$ (MeV)   &  1008.00    &  795.359 &  916.502 &   816.508   \\
$m_{\rho}$ (MeV)     &  763.000    &  763.000 &  763.000 &   763.000   \\
${\rm g}_{\sigma}$   &  12.3342    &  10.1377 &  9.9362  &   7.1031    \\
${\rm g}_{\omega}$   &  17.6188    &  13.2846 &  11.8188 &   8.8496    \\
${\rm g}_{\rho}$     &  10.3782    &  9.9514  &  10.0254 &   10.2070   \\
$b$ (fm$^{-1}$)      &  0.0        &  24.3448 &  0.0     &  24.0870    \\
$c$                  &  0.0        &$-$217.5876 &  0.0     & $-$15.9936   \\
\\
$\rho_{0}$ (fm$^{-3}$)& 0.1525     &  0.1518  &  0.1513  &   0.1524    \\
$E/A$ (MeV)           & $-$17.03   &  $-$16.43& $-$17.39 &  $-$16.98   \\
$m^{*}/M_{N}$         & 0.533      &  0.572   & 0.725    &   0.788     \\
$K$ (MeV)             & 580        &  212     &  480     &   294       \\
$a_{4}$ (MeV)         & 46.8       &  43.6    &  40.4    &   40.4      \\
\\
$\delta\rho$ in $^{208}{\rm Pb}$ (fm$^{-3}$)&
                        $-$0.0075 & $-$0.0070 & $-$0.0016&  $-$0.0030  \\
$\chi ^{2}$              &   1773   &   66      &  1040   & 812   \\      
\hline
\hline
\end{tabular}
\end{table}

\newpage
\begin{table}
\caption{The single-particle energies of both protons and anti-protons 
 as well as the binding energy per nucleon and the {\em rms} charge
radius in $^{16}{\rm O}$, $^{40}{\rm Ca}$ and
 $^{208}{\rm Pb}$. } \vspace{0.5cm}
\begin{tabular}{cccccc}
\hline
\hline
 & \multicolumn{2}{c}{RMF} & \multicolumn{2}{c}{RHA} &   \\
 & LIN & NL1 & RHA0 & RHA1 & EXP. \\
 \hline
$\;\;\;\;\;\;\;\;^{16}{\rm O}$ &       &          &           &      &       \\
$E/A$ (MeV)   &   7.80   &   8.00   &    8.01   & 8.00 &   7.98 \\
$r_{ch}$ (fm) &   2.59   &   2.73   &    2.62   & 2.66 &   2.74 \\
PROTONS     &          &          &           &      &        \\
$1s_{1/2}$ (MeV)& 42.99  &  36.18   &   32.21   & 30.68&   40$\pm$8 \\
$1p_{3/2}$ (MeV)& 20.71  &  17.31   &   16.09   & 15.23&   18.4   \\
$1p_{1/2}$ (MeV)& 10.85  &  11.32   &   12.98   & 13.24&   12.1  \\
ANTI-PRO.     &          &          &           &      &        \\
$1\bar{s}_{1/2}$ (MeV)& 821.30 &  674.11  &   413.62  &299.42&            \\
$1\bar{p}_{3/2}$ (MeV)& 754.62 &  604.70  &   369.78  &258.40&          \\
$1\bar{p}_{1/2}$ (MeV)& 755.43 &  605.77  &   370.36  &258.93&         \\
\hline
$\;\;\;\;\;\;\;^{40}{\rm Ca}$ &       &          &           &      &       \\
$E/A$ (MeV)   &   8.38   &   8.58   &    8.65   & 8.73 &   8.55 \\
$r_{ch}$ (fm) &   3.36   &   3.48   &    3.39   & 3.42 &   3.45 \\
PROTONS     &          &          &           &      &        \\
$1s_{1/2}$ (MeV)& 51.21  &  46.86   &   38.64   & 36.58&  50$\pm$11\\
$1p_{3/2}$ (MeV)& 35.05  &  30.15   &   27.11   & 25.32&          \\
$1p_{1/2}$ (MeV)& 29.25  &  25.11   &   25.17   & 24.03&  34$\pm$6  \\
ANTI-PRO.     &          &          &           &      &        \\
$1\bar{s}_{1/2}$ (MeV)& 840.76 &  796.09  &   456.58  &339.83&            \\
$1\bar{p}_{3/2}$ (MeV)& 792.36 &  706.36  &   424.85  &309.24&          \\
$1\bar{p}_{1/2}$ (MeV)& 792.75 &  707.86  &   425.14  &309.52&         \\
\hline
$\;\;\;\;\;\;\;^{208}{\rm Pb}$ &       &          &           &      &       \\
$E/A$ (MeV)   &   7.83   &   7.89   &    7.96   & 7.93 &   7.87 \\
$r_{ch}$ (fm) &   5.34   &   5.52   &    5.43   & 5.49 &   5.50 \\
PROTONS     &          &          &           &      &        \\
$1s_{1/2}$ (MeV)& 58.71  &  50.41   &   44.43   &40.80 &           \\
$1p_{3/2}$ (MeV)& 52.74  &  44.45   &   39.87   &36.45 &          \\
$1p_{1/2}$ (MeV)& 51.83  &  43.75   &   39.49   &36.21 &            \\
ANTI-PRO.     &          &          &           &      &        \\
$1\bar{s}_{1/2}$ (MeV)& 830.16 &  717.01  &   476.61  &354.18&            \\
$1\bar{p}_{3/2}$ (MeV)& 819.15 &  705.20  &   466.08  &344.48&          \\
$1\bar{p}_{1/2}$ (MeV)& 819.22 &  705.28  &   466.13  &344.52&   \\      
\hline
\hline
\end{tabular}
\end{table} 

\end{document}